\documentclass[prd,twocolumn,a4paper,floatfix,nofootinbib,preprintnumbers,superscriptaddress]{revtex4-2}
\usepackage[utf8]{inputenc}
\usepackage{hyperref}
\usepackage{xcolor}
\usepackage{braket}
\usepackage[nolist,nohyperlinks]{acronym}
\usepackage{amsthm}
\usepackage{graphicx}
\usepackage{amsfonts}
\usepackage{booktabs}
\usepackage{siunitx}
\usepackage{float}
\usepackage{mathtools}
\usepackage{multirow}
\usepackage{makecell}
\usepackage[normalem]{ulem}
\usepackage{orcidlink}
\usepackage{natbib}
\usepackage{bm}
\usepackage{cleveref}
\usepackage{xcolor}
\usepackage{caption}
\usepackage{subcaption}
\usepackage{placeins}
\usepackage[justification=RaggedRight]{caption}
\usepackage{transparent}

\usepackage{import}

\newcommand*\aap{A\&A}

\renewcommand*\ao{Appl.~Opt.}

\renewcommand*\apj{ApJ}
\newcommand*\apjl{ApJ}

\newcommand*\apjs{ApJS}

\newcommand*\mnras{MNRAS}

\renewcommand*\prd{Phys.~Rev.~D}

\AtBeginDocument{\RenewCommandCopy\qty\SI} 
\newcommand{\msun}{\text{M}_\odot}
\newcommand{\mtot}{M_{\text{tot}}}
\newcommand{\nsat}{\textit{n}_{\text{sat}}}
\newcommand{\chiEFT}{$\chi$EFT}
\newcommand{\mtov}{M_{\rm {TOV}}}
\newcommand{\fmiq}{\text{fm}^{-3}}

\newcommand{\sat}{\mathrm{sat}}
\newcommand{\sym}{\mathrm{sym}}

\graphicspath{{./}{Figures/}}
\usepackage{preamble}
\begin{document}

%\preprint{LA-UR-24-25688}
%\preprint{INT-PUB-24-026}

\title{Probe %Predict 
and Prejudice:\\ Classification of compact objects and model comparison using EOS knowledge}  

\author{Hauke \surname{Koehn}\,\orcidlink{0009-0001-5350-7468}}
\email{hauke.koehn@uni-potsdam.de}
\affiliation{Institut f\"ur Physik und Astronomie, Universit\"at Potsdam, Haus 28, Karl-Liebknecht-Str. 24/25, 14476, Potsdam, Germany}

\author{Thibeau Wouters\, \orcidlink{0009-0006-2797-3808}}
\email{t.r.i.wouters@uu.nl}
\affiliation{Institute for Gravitational and Subatomic Physics (GRASP), Utrecht University, Princetonplein 1, 3584 CC Utrecht, The Netherlands}
\affiliation{Nikhef, Science Park 105, 1098 XG Amsterdam, The Netherlands}

\author{Henrik \surname{Rose}\,\orcidlink{0009-0009-2025-8256}}
\email{henrik.rose@uni-potsdam.de}
\affiliation{Institut f\"ur Physik und Astronomie, Universit\"at Potsdam, Haus 28, Karl-Liebknecht-Str. 24/25, 14476, Potsdam, Germany}

\author{Peter T. H. \surname{Pang}\,\orcidlink{0000-0001-7041-3239}}
\affiliation{Institute for Gravitational and Subatomic Physics (GRASP), Utrecht University, Princetonplein 1, 3584 CC Utrecht, The Netherlands}
\affiliation{Nikhef, Science Park 105, 1098 XG Amsterdam, The Netherlands}

\author{Rahul \surname{Somasundaram}\,\orcidlink{0000-0003-0427-3893}}
\affiliation{Department of Physics, Syracuse University, Syracuse, New York 13244, USA}

\author{Ingo \surname{Tews}\,\orcidlink{0000-0003-2656-6355}}
\affiliation{Theoretical Division, Los Alamos National Laboratory, Los Alamos, New Mexico 87545, USA}

\author{Tim \surname{Dietrich}\,\orcidlink{0000-0003-2374-307X}}
\affiliation{Institut f\"ur Physik und Astronomie, Universit\"at Potsdam, Haus 28, Karl-Liebknecht-Str. 24/25, 14476, Potsdam, Germany}
\affiliation{Max Planck Institute for Gravitational Physics (Albert Einstein Institute), Am M{\"u}hlenberg 1, Potsdam 14476, Germany}

\begin{abstract}
\noindent 
Nuclear theory and experiments, alongside astrophysical observations, constrain the equation of state (EOS) of supranuclear-dense matter. 
Conversely, knowledge of the EOS allows an improved interpretation of nuclear or astrophysical data. 
In this article, we use several established constraints on the EOS and the new NICER measurement of \mbox{PSR J0437-4715} to comment on the nature of the primary companion in GW230529 and the companion of \mbox{PSR J0514-4002E}.
We find that, with a probability of $\gtrsim 84\%$ and $\gtrsim 68\%$, respectively, both objects are black holes.
These likelihoods increase to above 95\% when one uses GW170817's remnant as an upper limit on the TOV mass. 
We also demonstrate that the current knowledge of the EOS substantially disfavors high masses and radii for \mbox{PSR J0030+0451}, inferred recently when combining NICER with XMM-Newton background data and using particular hot-spot models. 
Finally, we also use our obtained EOS knowledge to comment on measurements of the nuclear symmetry energy, finding that the large value predicted by the PREX-II measurement displays some mild tension with other constraints on the EOS.
\end{abstract}

\maketitle

%\red{TO DO: Mention this work at some appropriate place down below} The GW230529 and PSR J0514-4002E were used to constrain the vector and isovector interactions in \ac{NS} matter in Ref.~\cite{Malik:2024qjw}. 

%%%%%%%%%%%%%%%%%%%%%%%%%%%%%%%%%%%%%%%%%%%%%%%%
\section{Introduction} \label{sec:intro}
%%%%%%%%%%%%%%%%%%%%%%%%%%%%%%%%%%%%%%%%%%%%%%%%

The dense-matter \ac{EOS} is a fundamental relation connecting nuclear physics in terrestrial experiments to astrophysical observations of \acp{NS}.
In describing these systems, the EOS spans a wide range in pressure and density.
Different research communities try to explore the EOS under various conditions through various physical messengers.
For example, nuclear physicists employ experiments and theory to constrain the EOS at nuclear densities, while astrophysicists focus on \ac{EOS} probes in \acp{NS} at the highest densities observable in the Universe.
We can then combine these data to place tighter constraints on the EOS.
Following this idea, a growing number of research articles, e.g., Refs.~\cite{Bauswein:2017vtn,Radice:2018ozg,Coughlin:2018fis,Capano:2019eae,Miller:2019nzo,Raaijmakers:2019dks,Dietrich:2020efo, Essick:2021kjb, Huth:2021bsp, Pang:2022rzc, Ghosh:2022lam, Essick:2021ezp, Takatsy:2023xzf}, have performed multimessenger studies including nuclear-physics computations. 
In the same spirit, we have previously compiled a comprehensive overview of available constraints on the dense-matter EOS~\citep{Koehn:2024set}.

%Neutron stars (NSs) are prime candidates for exploring the interplay of all fundamental interactions: gravity and electromagnetism, as well as weak and strong nuclear forces.
%Integrated studies of their formation, dynamics, interactions, and matter content can bridge gaps between nuclear theory, laboratory experiments, astronomical observations, and cosmological modeling, e.g., Refs.~\cite{Eichler:1989ve,Li:1998bw,Lattimer:2004pg,Lattimer:2006xb,Oertel:2016bki,Baym:2017whm}.
%In particular, NSs enable us to probe matter under the most extreme conditions in the Universe at densities several times the nuclear saturation density~\cite{Ozel:2016oaf,Miller:2019nzo,Burgio:2021vgk,Lattimer:2021emm}.
%Clearly, this makes NSs great astrophysical laboratories to test nuclear-physics principles and complement our knowledge obtained from nuclear-physics computations and experiments. 
%
Data gathered on NSs through, e.g., radio or X-ray observations of pulsars or through possible multimessenger detections of binary neutron star (BNS) systems, will become more and more precise, due to continuous upgrades of observational facilities and improved observational strategies.
As a consequence, systematic uncertainties will become increasingly important and need to be assessed and quantified when possible. 
Similarly, statistical or systematic outliers could arise in nuclear physics experiments. 
More precise constraints will allow important consistency checks of individual probes, allowing us to detect outliers.
These outliers could imply underestimated uncertainties or the existence of multiple branches of compact stars.
%Hence, comparing data points related to the EOS with each other is an important consistency check. 
Moreover, such comparisons also enable better interpretation of individual data points when additional EOS knowledge is included. 
For example, analyzing X-ray pulse profiles using different hot-spot configurations might describe the data comparably well but lead to different constraints on the EOS. 
In this case, considering EOS information from other sources might indicate a preference for one configuration over the other~\citep{Breschi:2024qlc, Miao:2024vka}.
Also, knowledge of the EOS can be used to classify newly observed compact objects~\citep{Golomb:2024mmt, Essick:2020ghc, Godzieba:2020tjn, Tews:2020ylw, LIGOScientific:2020aai, LIGOScientific:2020zkf}.

In the present article, we follow these ideas and revisit some recent measurements, showcasing how the Nuclear Physics Multi-Messenger Astrophysics framework NMMA~\cite{Dietrich:2020efo,Pang:2022rzc}\footnote{\url{https://github.com/nuclear-multimessenger-astronomy/nmma}} can contribute to multimessenger science. 
We provide updated constraints on the NS EOS by incorporating the most recent \ac{NICER} measurement of PSR J0437-04715~\cite{Choudhury:2024xbk} and use these constraints to comment on the plausibility of other recent data points on the EOS.
In particular, we use our updated EOS knowledge to classify the nature of two compact objects that fall in the putative lower mass gap, namely the primary component in GW230529~\cite{LIGOScientific:2024elc} and the companion of PSR J0514-04002E~\cite{Barr:2024wwl}.
We also determine model odds ratios for two data points that appear to be in slight tension with previous studies, namely the different models used to reanalyze \mbox{PSR J0030+0451} and the different measurements of the symmetry energy from nuclear experimental data~\cite{PREX:2021umo,CREX:2022kgg, Tamii:2011pv}.

The presentation of the established EOS constraints and the new NICER measurement is provided in Sec.~\ref{sec:EOS_constraints}, the potential mass gap objects are discussed in Sec.~\ref{sec:source_classification}, and the model comparison for potential outliers is given in Sec.~\ref{sec:model_selection}.

%%%%%%%%%%%%%%%%%%%%%%%%%%%%%%%%%%%%%%%%%%%%%%%%
\section{Available constraints on the Equation of State}
\label{sec:EOS_constraints}
%%%%%%%%%%%%%%%%%%%%%%%%%%%%%%%%%%%%%%%%%%%%%%%%

We can categorize current constraints on the NS EOS by the density range they explore.
Nuclear theory and experiments restrict the EOS at nuclear densities, around and slightly above the nuclear saturation density ${\nsat = \qty{0.16}{fm^{-3}}}$.
At higher densities, mass and radius measurements of stable NSs through electromagnetic observations as well as 
\ac{GW} and multimessenger signals of BNS or black-hole--neutron-star (BHNS) mergers provide additional means to explore the EOS~\cite{Koehn:2024set}.
However, the nature, model dependency, and constraining power of these different information channels vary drastically.

In this section, we first review well-known constraints that have an established standing in the literature and then look at the new mass-radius measurement for \mbox{PSR J0437-4715}~\cite{Choudhury:2024xbk}.

%%%%%%%%%%%%%%%%%%%%%%%%%%%%%%%%%%%%%%%%%%%%%%%%
\subsection{Established constraints}

In a previous study~\cite{Koehn:2024set}, we combined a variety of constraints to obtain a posterior estimate $P(\text{EOS}|d)$ on a set of 100\,000 model-agnostic EOS candidates.
Here, $d$ can refer to different sets of constraints (or `data') that we employed to infer the EOS. 
The EOS candidates were created with the nucleonic metamodel \citep{Margueron:2017eqc, Margueron:2017lup} in the regime around $\sim\qty{1}{\nsat}$ and extrapolated to higher densities with the speed-of-sound model.
A fixed crust EOS, taken from Ref.~\citep{Douchin:2001sv}, is attached to the metamodel at \qty{0.076}{\fmiq}.
Variations in the EOS of the crust may impact radii of lighter neutron star up to the order of \qty{0.5}{km}~\cite{Gamba:2019kwu, Grams:2021lzx}, but it seems to have little influence on the properties of higher mass neutron stars~\citep{Ferreira:2020zzy, Rather:2020gja}.
Hence, by keeping the crust EOS fixed, we might influence the results for the NICER measurements and the symmetry energy, though data uncertainties appear still wide enough to compensate for this effect. The compact object classification in Sec.~\ref{sec:source_classification} is likely not tarnished.
See Ref.~\citep{Koehn:2024set} for more details about the construction of our EOS candidate set.

\begin{figure}
    \centering
    \includegraphics[width = \linewidth]{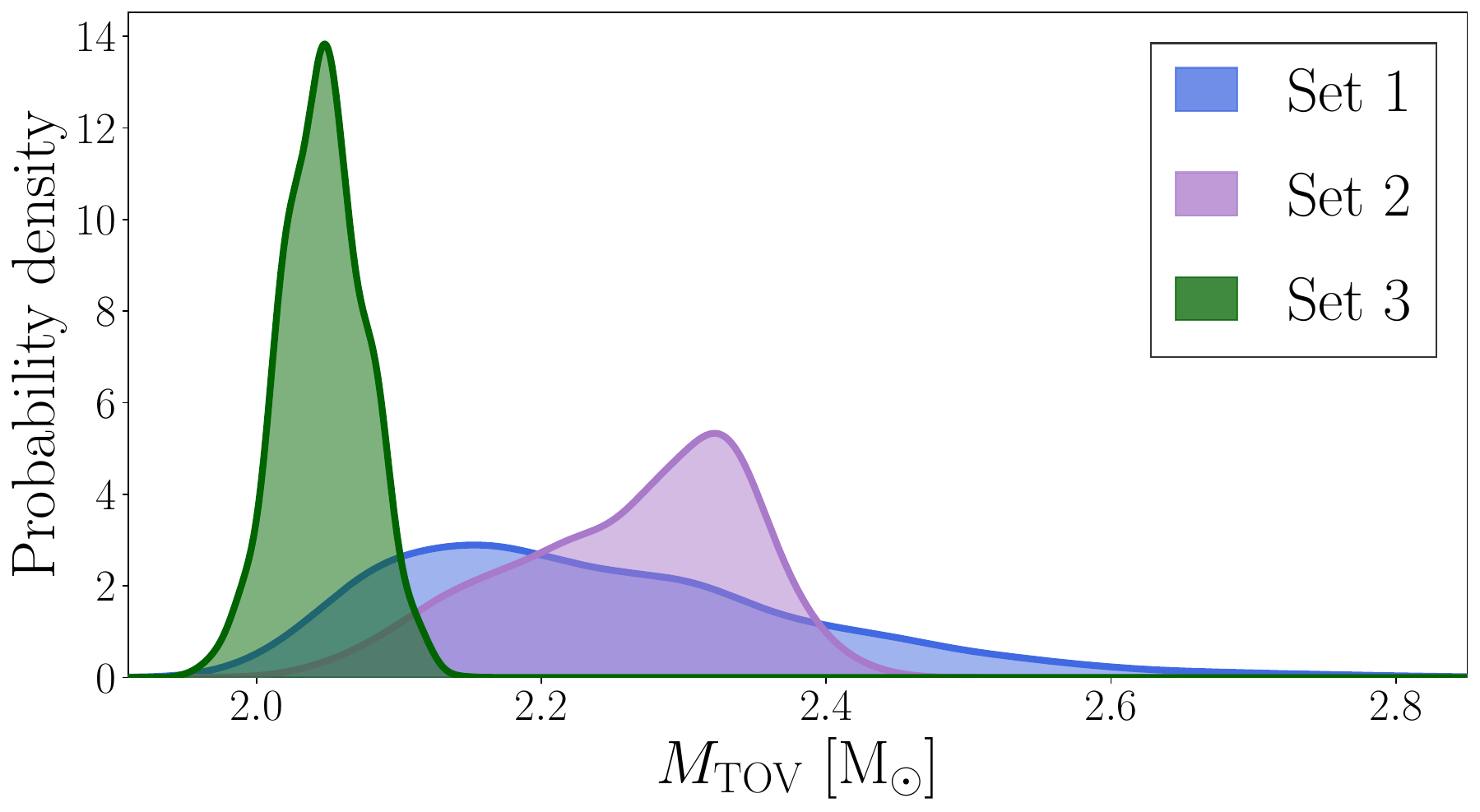}
    \caption{Kernel density estimates of the posterior distributions on the TOV mass $M_{\rm{TOV}}$ for the different sets of \ac{EOS} constraints given in Table~\ref{tab:sets}.}
    \label{fig: MTOV posterior distributions}
\end{figure}

For the present article, we use the EOS posterior estimates from Ref.~\cite{Koehn:2024set}. 
Since the constraining power and reliability of certain data points vary, we create three different sets of constraints as summarized in Table~\ref{tab:sets}.
Set 1 includes information from the radio timing observations of heavy pulsars, nuclear theory, the \ac{NICER} observations, and GW signals. 
In our judgement, these are the most widely accepted and model-independent constraints available; therefore, we characterize this set as ``high confidence''.
Set 2 is more vigorous as it makes use of additional constraints from Ref.~\cite{Koehn:2024set} that provide tighter restrictions on the EOS but introduce potential biases and are overall more model dependent. 
Among others, it includes an upper limit on the \ac{TOV} mass $\mtov$ (the maximum mass of a nonspinning \ac{NS}) that is derived from the assumption that the remnant of GW170817 collapsed to a \ac{BH}. 
The remnant mass can be inferred from GW and electromagnetic observations and then serves as an upper limit for $\mtov$. 
However, there is cautious evidence indicating that the GW170817 remnant was a short-lived hypermassive NS~\cite{Margalit:2017dij, Gill:2019bvq, Murguia-Berthier:2020tfs}. 
In that case, the remnant's mass would even yield an upper limit on the mass of a maximally rotating NS, i.e., the Kepler mass $M_{\text{Kep}}$, reducing $\mtov$ further. 
To also include this possibility in our analysis of the putative mass gap objects, we create an ``aggressive'' set 3, which encompasses the same constraints as set 2, but uses the GW170817 remnant mass as an upper limit on $M_{\text{Kep}}$.
The $\mtov$ posterior distributions for the different sets of constraints are shown in Fig.~~\ref{fig: MTOV posterior distributions}.
For further details, we refer to Ref.~\cite{Koehn:2024set}.

\begin{table}[t]
\renewcommand{\arraystretch}{1.1}
\caption{Overview of the EOS constraints contained within the three different datasets.}
\label{tab:sets}
\begin{tabular}{ p {1.7 cm}   p { 2 cm}  >{\raggedright\arraybackslash} p { 3.9cm}  }
\toprule
\toprule
Set & Label & Description \\
\midrule
\multirow{7}{8.5pt}{High confidence} & \multirow{7}{8.5pt}{Set~1} & \chiEFT, \linebreak pQCD, heavy~radio~pulsars, NICER~\mbox{J0740+6620}, NICER~\mbox{J0030+0451}, NICER~\mbox{J0437-4715}, GW170817\\
\midrule
\multirow{9}{8.5pt}{More vigorous} & \multirow{9}{8.5pt}{Set~2} & Set 1,\linebreak PSR~\mbox{J0952-0607}, heavy~ion-collisions, GW170817+KN+GRB,\linebreak GW170817 remnant mass, \linebreak qLMXBs in $\omega$ Centauri and 47 Tucanae,\linebreak \mbox{4U 1702-429}, SAX~\mbox{J1808.8-3658}\\
\midrule
\multirow{4}{8.5pt}{Aggressive} & \multirow{4}{8.5pt}{Set~3} & Same as set 2, but for the remnant of GW170817 a hypermassive NS above the Kepler limit is assumed\\
\bottomrule
\end{tabular}
\end{table}
%%%%%%%%%%%%%%%%%%%%%%%%%%%%%%%%%%%%%%%%%%%%%%%%
\subsection{Recent NICER observation of PSR J0437-4715}
\label{sec:new_NICER}

\begin{table}
\caption{Credible intervals on EOS properties when different constraints are employed. Limits are shown at 95\% credibility. The upper row shows the limit before  \mbox{PSR J0437-4715} is included, the lower row lists them afterward. The bracketed values show the Kullback-Leiber divergences between the two. $R_{1.4}$ is the radius of a canonical neutron star, $\mtov$ the TOV mass, $p_{3\nsat}$ the pressure at three times saturation density, and $n_{\text{TOV}}$ the central density for a TOV star.}
\label{tab:new_NICER}
\renewcommand{\arraystretch}{1.25}
\begin{tabular}{>{\arraybackslash} p {1.5 cm}  >{\centering\arraybackslash} p { 2 cm}  >{\centering\arraybackslash} p { 1.5 cm}  >{\centering\arraybackslash} p { 1.5 cm} >{\centering\arraybackslash} p { 1.5 cm}  }
 \toprule
 \toprule
 Quantity & \mbox{PSR J0437} included & Set 1 & Set 2 & Set 3\\
 \midrule
\multirow{3}{8.5pt}{$R_{1.4}$~[km]} & $\cross$ & $12.26^{+0.80}_{-0.91}$& $12.39^{+0.48}_{-0.54}$& $12.40^{+0.50}_{-0.59}$\\
& \checkmark  & $12.05^{+0.84}_{-0.82}$& $12.33^{+0.48}_{-0.55}$& $12.33^{+0.50}_{-0.62}$\\
&  & (0.092)  & (0.029)  & (0.034) \\
\midrule\multirow{3}{8.5pt}{$M_{\text{TOV}}$~[M$_\odot$]} & $\cross$ & $2.25^{+0.42}_{-0.22}$& $2.28^{+0.11}_{-0.20}$& $2.05^{+0.06}_{-0.06}$\\
& \checkmark  & $2.22^{+0.38}_{-0.19}$& $2.28^{+0.11}_{-0.20}$& $2.05^{+0.06}_{-0.06}$\\
&  & (0.020)  & (0.001)  & (0.001) \\
\midrule\multirow{3}{8.5pt}{$p_{3 n_{\text{sat}}}$ [MeV~fm$^{-3}$]} & $\cross$ & $90^{+71}_{-31}$& $94^{+32}_{-21}$& $75^{+17}_{-12}$\\
& \checkmark  & $84^{+64}_{-28}$& $93^{+32}_{-22}$& $74^{+19}_{-13}$\\
&  & (0.048)  & (0.004)  & (0.013) \\
\midrule\multirow{3}{8.5pt}{$n_{\text{TOV}}$~[$n_{\text{sat}}$]} & $\cross$ & $5.92^{+1.34}_{-1.38}$& $5.85^{+0.91}_{-0.92}$& $6.76^{+0.71}_{-1.09}$\\
& \checkmark  & $6.12^{+1.29}_{-1.39}$& $5.86^{+0.94}_{-0.92}$& $6.82^{+0.70}_{-1.11}$\\
&  & (0.039)  & (0.001)  & (0.005) \\
\bottomrule
\end{tabular}
\end{table} 

\ac{NICER} measures high-time-resolution X-ray spectra and is therefore able to track X-ray pulses from the surface of NSs. 
Modeling these pulses allows for the extraction of a posterior distribution $P(M, R|\text{NICER})$ on the mass and radius of the NS. 
Two NSs have been investigated using this method before~\cite{Riley:2019yda, Miller:2019cac, Riley:2021pdl, Miller:2021qha}. 
More recently, a new measurement of \mbox{PSR J0437-4715} (herein abbreviated \mbox{PSR J0437}) has become available~\cite{Choudhury:2024xbk}. 
Together with the mass measurement from radio timing data, the most successful hot-spot model (CST+PDT) assumed one emitting concentric single-temperature hot-spot ring as well as a dual-temperature spherical hot-spot and inferred a mass of $M=1.42_{-0.07}^{+0.07}\,\msun$ and $R = 11.36_{-1.15}^{+1.91}$\,km (95\% credibility). 

However, the analysis of PSR~J0437 was complicated by the presence of a time-variable active galactic nucleus (AGN) near the source, making the background estimation more challenging.
In particular, the headline result uses constraints on the background estimate from the 3C50 model~\cite{Remillard:2021cjp} and time-averaged NICER spectra of the AGN. 
If this background constraint is not applied, the inferred radius is $11.01^{+0.51}_{-0.48}$\,km (68\% credibility) without any distribution tail toward higher values. 
Complementary analysis with time-averaged XMM-Newton data confirmed the consistency of the background constraints, but if the time variability of the AGN biases the background estimation, then this might lead to differences in the inferred radii.
Similarly, one of the two emitting hot-spot components is likely relatively broad and noncircular in shape, and thus it seems possible that different hot-spot geometries than CST+PDT could also fit the data well while resulting in slightly different radius estimates.
Moreover, despite being above NICER's actual frequency range~\citep{Choudhury:2024xbk}, the detection of a nonthermal pulsed component above \qty{3}{keV} with NuSTAR could, in principle, also impede the results.

Unlike previous NICER measurements, the result also lacks confirmation from an independent second analysis.
Therefore, it is worth keeping in mind that potential systematic biases in the analysis of the new NICER measurement could lead to offsets in the estimated radius.
For all of these reasons, in the Appendix we also present how the results of our subsequent analyses in Secs.~\ref{sec:source_classification} and~\ref{sec:model_selection} change, when we exclude PSR~J0437 from the constraint sets.

When applying the new NICER measurement of PSR~J0437, we use the posterior of the headline result in Ref.~\citep{Choudhury:2024xbk} to further constrain the EOS set of Ref.~\cite{Koehn:2024set} with the following likelihood:
\begin{align}
\begin{split}
    &\mathcal{L}(\text{EOS}|\text{PSR J0437}) =\\
    &\int_0^{M_{\text{TOV}}} \differential M\ P(M, R(M, \text{EOS})| \text{PSR J0437})\,.
\end{split}
\label{eq:NICER_likelihood}
\end{align}
The samples for the posterior $P(M, R| \text{PSR J0437})$ are taken from Ref.~\citep{Choudhury_2024_samp}.

In Table~\ref{tab:new_NICER}, we report how the posterior estimates on EOS-derived quantities change when this new constraint is added to sets~1, 2, and 3 from Table~\ref{tab:sets}. 
Naturally, the impact of the new NICER measurement is largest for set~1, given that sets~2 and 3 already encompass a larger collection of other constraints. 
In any case, \mbox{PSR J0437} affects mostly the radii of intermediate-mass \acp{NS} and has only little impact on $\mtov$. 
Generally, the posterior is shifted toward slightly softer EOS, given the comparatively small radius inferred by NICER.

To compare the new mass-radius measurement of \mbox{PSR J0437} with respect to the expectation from the previous constraints, we employ the technique of posterior predictive checking. 
Having a posterior $P(\text{EOS}|d)$ from some constraint set, we can calculate the posterior predictive distribution for the \mbox{PSR J0437} inference as 
\begin{align}
    &P(\text{PSR J0437}|d)= \sum_{\text{EOS}} \mathcal{L}(\text{EOS}|\text{PSR J0437}) P(\text{EOS}|d)\,.
    \label{eq:posterior_predictive}
\end{align}
Since the absolute value of this expression is not of particular interest, we instead compare it to the radius estimate we would get directly from $P(\text{EOS}|d)$. 
Accordingly, we predict a mock \mbox{PSR J0437} posterior, denoted $P(M, R|\text{PSR J0437*})$, that uses the same mass samples as the original measurement, but the radii are drawn using $P(\text{EOS}|d)$. 
For that posterior, we also determine the likelihoods according to Eq.~\eqref{eq:NICER_likelihood} and calculate the ratio of the posterior predictiveness between the real NICER measurement and our mock radius prediction. 
In that way, we compare the plausibility of different $M$-$R$ posteriors for \mbox{PSR J0437}. 
The posterior predictiveness ratio works in a manner similar to the usual Bayes factor. 

We find that before we add \mbox{PSR J0437} to sets~1, 2, and 3 the respective EOS posteriors predict a radius of $12.27^{+0.80}_{-0.91}$, $12.39^{+0.47}_{-0.54}$, and $12.40^{+0.50}_{-0.59}$\,km for \mbox{PSR J0437}.
This translates to a posterior predictiveness ratio to the real measurement of 2.42, 5.17, and 5.00, respectively.
Hence, the hitherto expectation about the EOS would predict a radius of $\approx 12.3$\,km, and such a value would be two to five times more plausible than the actual measurement yields. 
If we excluded all the other NICER measurements from set 1, we would expect a radius of ${11.98}^{+1.09}_{-1.08}$\,km for PSR~J0437, yielding a posterior predictiveness ratio of 1.53 to the actual measurement. 

Despite the possibility of a relatively small radius in the new NICER measurement, which would imply a shift toward softer EOS consistent with nuclear theory and GW observations, our statistical analysis does not indicate a significant tension with expectations given the present uncertainties. 
Both the predicted and the measured values for the radius of \mbox{PSR J0437} overlap in their $95\%$ credibility limits, and the posterior predictiveness ratios of $\lesssim 3$ for set 1 point toward the consistency between the new measurement and the established constraints from chiral effective field theory (\chiEFT), perturbative quantum chromodynamics (pQCD), the heavy pulsars, and GW170817. 
The constraints in sets 2 and 3, namely the large mass of the Black Widow \mbox{PSR J0952-0607} and the larger radius of \mbox{4U 1702-429}, favor slightly stiffer EOS, increasing the posterior predictive ratio to $\approx 5$, indicating some difference between expectation and measurement, though still at an acceptable level.

%%%%%%%%%%%%%%%%%%%%%%%%%%%%%%%%%%%%%%%%%%%%%%%%
\section{Classification of binary Components} \label{sec:source_classification}
%%%%%%%%%%%%%%%%%%%%%%%%%%%%%%%%%%%%%%%%%%%%%%%%
The TOV mass $\mtov$ is the highest possible mass for a nonspinning NS.
Its exact value depends on the EOS. 
Knowing $\mtov$ and ignoring effects due to rotation, one can decide whether a compact object with a given mass should be considered a \ac{BH} or \ac{NS}. 
Similarly to Ref.~\cite{Tews:2020ylw}, we apply the $\mtov$ distribution of our EOS sets to perform a Bayesian source classification for two recently reported compact objects, namely the companion of PSR J0514-4002E~\citep{Barr:2024wwl} and the primary of GW230529~\citep{LIGOScientific:2024elc}.
Their mass estimates fall into the posited lower mass gap.

Mass measurements on \acp{LMXB} have identified a remarkable lack of systems where the accreting compact object's mass lies in a range of about \qtyrange{2}{5}{\msun}.
Ref.~\cite{Bailyn:1998} first indicated this lower mass gap between massive \acp{NS} and low-mass \acp{BH}.
The existence of such a mass gap has been subject to various challenges since early on~\citep{Fryer:2001}, but has also received subsequent support by additional observations and supernova modeling efforts.e
For instance, Ref.~\cite{Ozel:2010} concluded from a larger sample of \ac{BH} binaries that a relatively narrow Gaussian distribution $\mathcal{N}(\mu=7.8 \msun,\ \sigma=1.2\msun)$ would best describe the \ac{BH} mass spectrum of LMXBs and that this could not be reconciled with a naive explanation of the lower mass gap as an observational selection effect.
Semi-analytical remnant-mass prescriptions for supernovae, therefore, explained the mass gap as an effect of the explosion timescale~\citep{Belczynski:2012,Fryer:2012}.
However, in Ref.~\cite{Janka:2012} it was pointed out that current supernova modeling lacks the robustness required to firmly establish this link.
Recently, simulations showed evidence that \acp{BH} in the lower mass gap can form in core-collapse supernovae, e.g., Ref.~\cite{Boccioli:2024kvw}.
Identifying fundamental properties of mass gap objects can provide highly informative constraints on compact object formation, on both the individual and population levels.

To classify the compact objects mentioned above, we use the posterior on $P(\mtov|d)$ derived from our constrained EOS set. 
If there is an observation $O$ of a compact object and its mass posterior is inferred to be $P(M|O)$, we can determine the probability of this compact object being a NS as~\cite{Tews:2020ylw}
\begin{align}\label{eq: probability of NS}
\begin{split}
    P(\text{NS}) &= \sum_{\text{EOS}} P(\text{EOS}|d) \int_{0}^{\mtov(\text{EOS})}\! \! \!  \differential M\ P(M|O) \\
     &= \!\int \!\differential\mtov \!\int_{0}^{\mtov}\! \! \! \differential M\ P(\mtov|d) P(M|O).
     % \\
     % &= \int_0^{\infty}\ \differential\Delta m \int_{-\infty}^{\infty} \differential m\ P(m+\Delta m|d) P(m|O).
\end{split}
\end{align}
However, the above integral is only valid for nonspinning NSs. 
To include information on additional support due to high dimensionless spins $\chi$, the maximal mass of a \ac{NS} can be higher, meaning $M_{\rm{max}}(M_{\rm TOV}, \chi)>M_{\rm TOV}$.
For our analysis of the primary component of GW230529, we use the quasiuniversal relation for $M_{\rm{max}}$ from Ref.~\cite{Breu:2016ufb} and extend the above probability estimate as
\begin{align}
\begin{split}
\label{eq: probability of NS with spin}
        &P(\text{NS}) =\\&\int \differential M_{\rm{max}}\int_{0}^{M_{\rm{max}}} \! \! \!\differential M\int_0^1 \differential \chi \ P(M_{\rm{max}}|d, \chi) P(M, \chi |O) \, .
\end{split}
\end{align}
In the classification of compact objects below, we compute these integrals with a Monte Carlo approach and report the mean values averaged over $10$ realizations.\footnote{The standard deviations are several orders of magnitude smaller than the means and will therefore not be quoted.} 
We emphasize that our analysis assumes a disjoint distribution of \ac{BH} and \ac{NS} masses, i.e., we do not consider the possibility that these objects were primordial \acp{BH} or represented some exotic twin-star configurations~\cite{Essick:2024}.

%%%%%%%%%%%%%%%%%%%%%%%%%%%%%%%%%%%%%%%%%%%%%%%%
\subsection{Pulsar timing observations of PSR J0514-4002E}

\mbox{PSR J0514-4002E} is a \ac{MSP} and was first identified in a survey of the globular cluster NGC~1851~\cite{Ridolfi:2022}.
Subsequent radio observations were reported in Ref.~\cite{Barr:2024wwl}.
Their timing solutions from the \textsc{TEMPO} code~\cite{Hobbs:2006cd} demonstrated compellingly that this pulsar lives in a binary system.
These timing solutions determined Keplerian orbital parameters to high precision too, obtaining the binary mass function as 
\begin{align}
f(M_1, M_2) & = \frac{(M_1\, \sin i)^3}{(M_2+M_1)^2} %\nonumber\\
            % & = 4\pi^2 \frac{c^3}{G} \frac{x^3}{P_b^2} 
            = \qty{0.41672(22)}{M_\odot}.
            \label{eq:mass_function}
\end{align}
For the sake of consistency throughout this paper, but in contrast to conventions in pulsar astronomy, we label the binary components as $M_1$ and $M_2$, where $M_1$ denotes the unseen massive object of interest and $M_2$ is the observed pulsar. 
% We here denote the binary component masses as $M_1>M_2$, i.e., in this case $M_1$ denotes the unseen massive object of interest and $M_2$ is the observed pulsar.
The inclination angle $i$ is defined such that $i = 0$ corresponds to a face-on orbit.
Among the post-Keplerian parameters, the advance of periastron $\Dot{\omega}$ was tightly constrained and provided an estimate for the total mass as 
\begin{align}
    M_{\text{tot}}=M_1+M_2=\qty{3.8870(45)}{\msun}\,.
    \label{eq:total_mass}
\end{align} 
Combining these constraints implies $M_1\geq \qty{1.84}{\msun}$.
The nondetection of an optical counterpart in dedicated follow-up observations thus established the unseen component as a compact object.

In order to obtain a consistent mass estimate for $M_1$, Ref.~\cite{Barr:2024wwl} subsequently conducted a Bayesian analysis.
The authors used $M_{\text{tot}}$ and $\cos i$ as sampling parameters because these --ignoring uncertainty in the value for the mass function Eq.~\eqref{eq:mass_function} --uniquely determine $M_1$ while being more directly linked to the observational data: 
$M_{\text{tot}}$ corresponds by Eq. \eqref{eq:total_mass} to the observable $\Dot{\omega}$ and a flat prior in $\cos i$ represents our prior ignorance of the orbital orientation.
Moreover, they enforced \qty{1.17}{\msun} as an absolute lower limit for the pulsar mass $M_2$, inspired by the tightly constrained companion mass of \mbox{PSR J0453+1559}~\citep{Martinez:2015}.
Ref.~\cite{Barr:2024wwl} then generated new timing solutions with sampled pairs of ($M_{\text{tot}}, \cos i$) values as fixed inputs and weighted these by a likelihood function based on the resulting residuals.
Their posterior for $M_1$ suggests a probability that the massive companion is an \ac{NS} of around $30\%$ under the EOS constraints of set 1. 
Employing the more aggressive constraints of set 3 reduces this probability to only $2.3\%$.

We note briefly that the posterior for $M_1$ in Ref.~\cite{Barr:2024wwl} can be reproduced through a greatly simplified analysis, leveraging their intermediate \textsc{TEMPO} results. 
To that end, we sample the mass function $f$ and the total mass $\mtot$ from Gaussian priors corresponding to the quoted uncertainties in Eqs.~(\ref{eq:mass_function}) and (\ref{eq:total_mass}), along with an isotropic prior for the inclination. 
The full timing solutions provide estimates for additional post-Keplerian parameters, in particular for
\begin{align}
\begin{split}
    \gamma_E &:= e \left(\frac{P_b G^2}{2\pi c^6} \right)^{1/3} \frac{M_1(\mtot+M_1)}{\mtot^{4/3}} \\
    &~= \qty{0.0111(84)}{\second}\,, \\%\nonumber \quad\text{ and }\\
\end{split}
\end{align}
and
\begin{align}
    h_3 &:= \frac{G M_1}{c^3} \left(\frac{\sin i}{1+\cos i} \right)^3= \qty{- 0.02(91)}{\micro\second} \, ,
\end{align}
which relate to the Einstein delay and Shapiro delay, respectively~\citep{Lorimer:2008,Freire:2010}. Here, $P_b$ is the orbital period of the binary.
Although \textsc{TEMPO} allows negative fits, $\gamma_E$ and $h_3$ are positive by definition.
We treat these values and the reported correlation coefficient as characteristics of a bivariate Gaussian likelihood function that we then apply to our sampled values of $f$, $\mtot$, and $\cos i$. 
With this approach, we are able to reproduce the $M_1$-posterior of Ref.~\cite{Barr:2024wwl} to high precision.

However, this result is clearly subject to prior choices and alternative choices appear equally justified.
For instance, there are various, yet more uncertain measurements of \acp{NS} less massive than \qty{1.17}{\msun}, e.g., Ref.~\citep{Ozel:2016oaf}. 
Binary evolution theory allows lower NS masses too~\citep{Tauris:2015}.
Therefore, to obtain a more conservative estimate for $M_1$, we test an agnostic approach that does not constrain $M_2$ to an allowed mass range. 
Under this assumption, the NS hypothesis loses further support, dropping below $30\%$ for sets 1 and 2  and below $2\%$ for set 3.

Additionally, we can obtain a potentially more realistic $M_1$ estimate when taking our knowledge of the \ac{MSP} population into account.
Canonically, very high pulsar spins result from angular momentum conservation during mass transfer episodes. 
These circularize the system during late stellar evolution of the companion which was initially less massive than the pulsar's progenitor.
The companion star then evolves in nonfringe scenarios into a less massive object than the previously formed pulsar.
As remarked in Ref.~\cite{Barr:2024wwl}, the observed eccentricity and mass ratio then suggest that \mbox{PSR J0514-4002E} acquired its massive companion in a dynamic exchange encounter.
Although this disallows detailed conclusions on the companion based on binary evolution models, it indicates an intermediate to high pulsar mass.
Ref.~\cite{Antoniadis:2016} parameterizes the galactic \ac{MSP} mass distribution as a bimodal Gaussian
\begin{align}
\begin{split}
    P^{\rm gal}(M_{\rm PSR} [\unit{\msun}]) =& 0.611\, \mathcal{N}(\mu=1.396, \sigma=0.045) \\ 
    +& 0.389\,\mathcal{N}(\mu=1.84, \sigma=0.10) \, .
\end{split}
\end{align}
If $M_2$ were to sit in the higher-mass peak, the binary would roughly be an equal-mass system.
The mass function could then only result from low inclination which is strongly disfavored by the nondetection of Shapiro delay.
This suggests that indeed the pulsar mass could match the narrow first peak around the canonical pulsar mass of $\qty{1.4}{\msun}$, and hence the companion object would have a mass of $M_1=\qty{2.48(5)}{\msun}$. 
Correspondingly, its probability to be a \ac{NS} drops to only $14\%$ when employing set 1 and to 1.3\% for set 3.

\begin{figure}
    \centering
    \includegraphics[width = \linewidth]{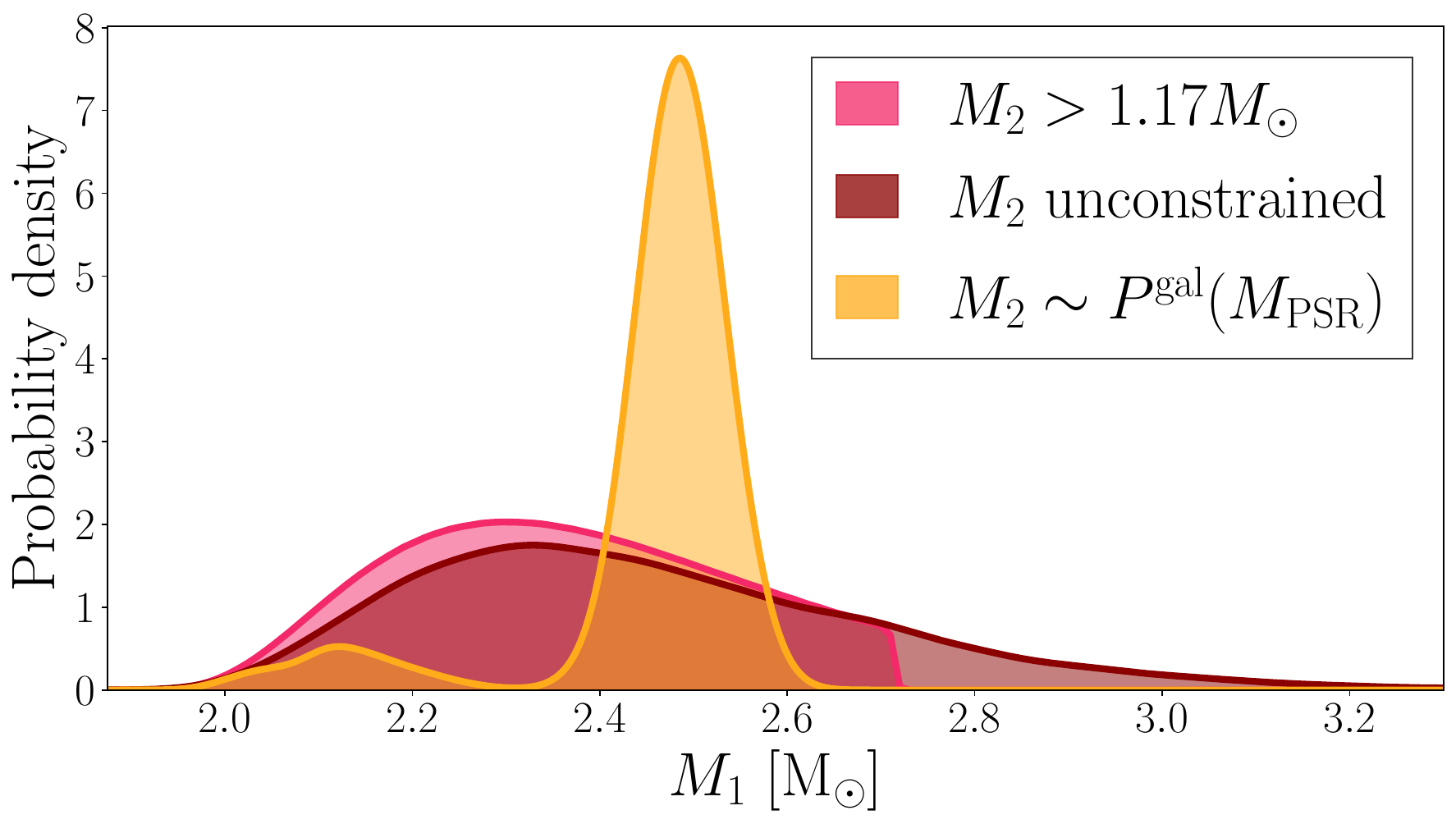}
    \caption{Posteriors on the mass of the companion of PSR J0514-4002E with different priors: constraining the mass of the pulsar to be above $1.17$ M$_\odot$ (pink), assuming an isotropic prior (dark red), and using a population-informed prior (orange).}
    \label{fig:PSR posteriors}
\end{figure}

The $M_1$-posterior distributions obtained under the various priors discussed above are shown in Fig.~\ref{fig:PSR posteriors}. The probabilities for the companion to be an \ac{NS} are summarized in Table~\ref{tab: classification results}.

\subsection{Gravitational-wave event GW230529}

On the 29th of May 2023 at 18:15:00 UTC, the LIGO Livingston observatory~\cite{LIGOScientific:2014pky} detected a compact binary merger signal, GW230529\_181500, simply referred to as GW230529 from here on.
All other \ac{GW} observatories were either offline or did not have the required sensitivity to detect the signal at the time of the merger. 
Nevertheless, the signal was reported by three matched-filter search pipelines, \textsc{GstLAL}~\cite{Messick:2016aqy, Sachdev:2019vvd, Hanna:2019ezx, CANNON2021100680, Ewing:2023qqe, Tsukada:2023edh}, \textsc{MBTA}~\cite{Adams:2015ulm, Aubin:2020goo}, and \textsc{PyCBC}~\cite{Allen:2005fk, Allen:2004gu, DalCanton:2020vpm, Usman:2015kfa, Nitz:2017svb, Davies:2020tsx}, with \acp{SNR} of $11.3$, $11.4$, and $11.6$, respectively, making an astrophysical origin of the source likely.
The masses of the primary and secondary components of GW230529's source were estimated to be $M_1 = 3.6^{+0.8}_{-1.2}$\,$\msun$ and $M_2 = 1.4^{+0.6}_{-0.2}\,\msun$, respectively, with uncertainties reported at the $90\%$ credible level. 
As such, the primary object belongs to the lower mass gap, which has triggered already a number of follow-up studies about the origin of this binary system, see, e.g., Refs.~\cite{Huang:2024wse,Chandra:2024ila,Zhu:2024cvt}. 
Moreover, GW230529 was already used to test general relativity (GR)~\cite{Sanger:2024axs} and constrain beyond-GR theories~\cite{Gao:2024rel}. 

The source properties were originally inferred from waveform models in Ref.~\cite{LIGOScientific:2024elc}, which lead to the conclusion that a \ac{NSBH} merger was the most viable explanation for the event.
To account for systematic errors in waveform modeling, which might be significant for \ac{NSBH} mergers~\cite{Huang:2020pba}, the final results of Ref.~\cite{LIGOScientific:2024elc} combined the posterior distributions obtained with the inspiral-merger-ringdown phenomenological waveform \texttt{IMRPhenomXPHM}~\cite{Pratten:2020fqn, Garcia-Quiros:2020qpx, Pratten:2020ceb} and the effective one-body model \texttt{SEOBNRv5PHM}~\cite{Khalil:2023kep, Pompili:2023tna, Ramos-Buades:2023ehm, vandeMeent:2023ols}, which we will refer to as the ``default'' set of posterior samples.  
While these waveforms are used for binary black hole waveforms, Ref.~\cite{LIGOScientific:2024elc} did not find evidence that BNS~\cite{Dietrich:2017aum,Dietrich:2019,Dietrich:2019kaq} or NSBH~\cite{Thompson:2020nei,Matas:2020wab} waveforms are preferred, due to the moderate \ac{SNR} of GW230529. 

The interpretation as a \ac{NSBH} leads to various insights regarding these systems.
For instance, GW230529's component masses are significantly more symmetric than previously observed \ac{NSBH} \ac{GW} events~\cite{KAGRA:2021vkt, LIGOScientific:2021qlt}, which raises the expected rate of \ac{NSBH} events with an accompanying electromagnetic counterpart.

Since the tidal information of GW230529 itself is inconclusive and no electromagnetic counterpart has been observed~\cite{Ahumada:2024qpr, Ronchini:2024lvb, Coulter:2024eml}, the nature of the primary object was constrained only based upon its inferred mass and spin~\cite{Essick:2020ghc}.
% Ref.~\citep{LIGOScientific:2024elc} determined the probability of the primary object being an \ac{NS} in a similar way to Eq.~\eqref{eq: probability of NS}. 
However, Ref.~\cite{LIGOScientific:2024elc} considered an \ac{EOS} parametrized using Gaussian processes~\cite{Landry:2018prl, Essick:2019ldf} and conditioned on observations of massive pulsars and previous \ac{GW} measurements~\cite{Landry:2020vaw}. 
In this work, on the other hand, we make use of the compiled sets of \ac{EOS} constraints from Table~\ref{tab:sets} and their corresponding posteriors on the \ac{TOV} mass used in Eqs.~\eqref{eq: probability of NS} and \eqref{eq: probability of NS with spin}. 

\begin{figure}
    \centering
    \includegraphics[width = \linewidth]{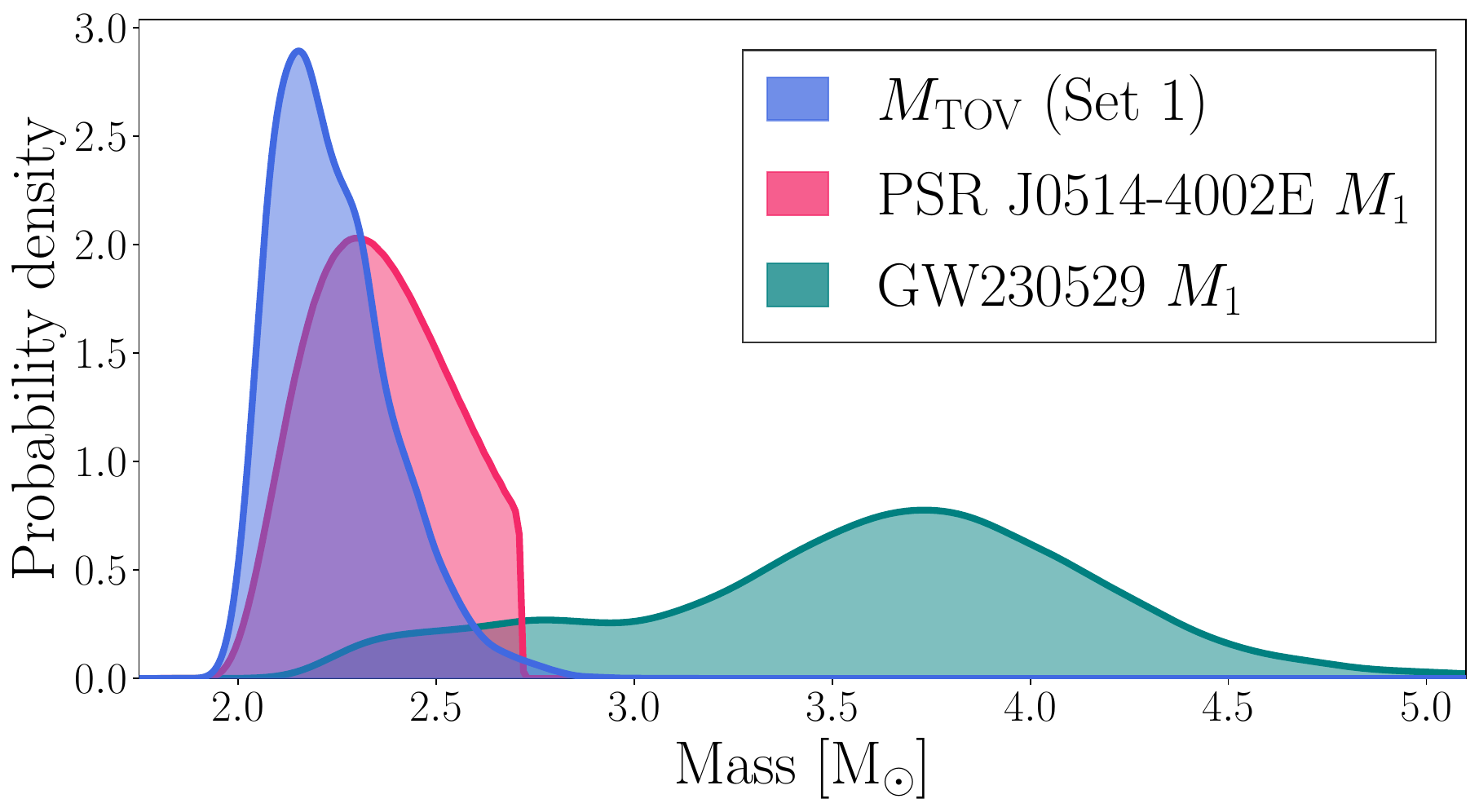}
    \caption{Visual comparison between the posterior distributions of $M_{\rm{TOV}}$ from set~1 (blue), the PSR J0514-4002E companion enforcing the $M_2 > 1.17$ M$_\odot$ constraint (pink) and the GW230529 primary object using the default priors (teal).}
    \label{fig: mass posteriors for classification}
\end{figure}

We consider two different posterior distributions of the source mass\footnote{We do not consider the possibility that the GW event is lensed, which can shift the mass distributions of the objects~\cite{Canevarolo:2024muf, Bianconi:2022etr}.} and spin of the primary object, both obtained from the public data release of GW230529~\cite{ligo_scientific_collaboration_2024_10845779}.
First, we use the default set of posterior samples, as described above. 
The mass distribution of the default posterior set is visually compared against the $M_{\rm{TOV}}$ posterior of set~1 in Fig.~\ref{fig: mass posteriors for classification}. 
Second, we also use the posterior samples obtained from reweighting the default posterior samples according to the \textsc{Power law + Dip + Break} (\textsc{PDB}) population model~\cite{Fishbach:2020ryj, Farah:2021qom}.
\textsc{PDB} models the mass distribution of binary components observed in \acp{GW} events with an ansatz made up of a power law, a dip to account for the lower mass gap, and a break in the power law behavior to allow the \ac{BH} and \ac{NS} regions to have distinct power law coefficients. 

In Table~\ref{tab: classification results}, we show the probabilities of the primary component of GW230529 being an \ac{NS} with and without taking the spin information into account, i.e., using Eqs.~\eqref{eq: probability of NS} \eqref{eq: probability of NS with spin}, respectively. 
As expected, incorporating the spin information of the primary object raises the probability of it being an \ac{NS}. 
We note that the probability decreases significantly once we consider the more vigorous and strict sets of EOS constraints. 
Therefore, we conclude that the primary object of GW230529 is most likely a \ac{BH}.

Our results agree with  Ref.~\cite{LIGOScientific:2024elc} which reports a $\sim 2.9\%$ ($\lesssim 0.1 \%$) probability for the primary being a NS when considering an astrophysics-agnostic high-spin (low-spin) prior, and $\sim 8.8 \%$ in case the \textsc{PDB} population prior is used. 
These analyses, together with our results, show that the probability of a compact object belonging to the lower mass gap being a neutron star strongly depends on the inclusion of spin information of the object, as well as the EOS constraints and priors used in the analysis.

\begin{table}
    \renewcommand{\arraystretch}{1.25}
    \centering
    \caption{Probability that the companion of \mbox{PSR J0514-4002E} and the primary component of GW230529 are NSs. The various assumptions used for the analysis of \mbox{PSR J0514-4002E} are discussed in the main text. The analysis of GW230529 considers the default priors (explained in the main text) and the priors informed by the \textsc{Power law + Dip + Break} (\textsc{PDB}) population model.}
    \label{tab: classification results}
    \begin{tabular*}{0.975\linewidth}{@{\extracolsep{\fill}} p {1.8 cm} p{1.2 cm} >{\centering\arraybackslash}p {1 cm}  c c c}
\toprule\toprule
Object & & & Set $1$ & Set $2$ & Set $3$ \\
\midrule % TODO: update final values here
\multirow{3}{*}{\makecell{PSR J0514\\-4002E}} & \multicolumn{2}{l}{$M_2>\qty{1.17}{\msun}$}& $29\%$ & $32\%$ & $\phantom{0}2.3\%$ \\ 
 & \multicolumn{2}{l}{$M_2$ unconstrained} & $23\%$ & $25\%$ & $\phantom{0}1.3\%$ \\ 
 & \multicolumn{2}{l}{$M_2\in P^{\rm gal}(M_{\rm PSR})$} & $13\%$ & $\phantom{0}7.8\%$ & $\phantom{0}1.2\%$ \\ 
 \midrule \multirow{8}{*}{GW230529} & \multirow{4}{*}{\makecell{default \\ prior}} & w/o spin & \multirow{2}{*}{$\phantom{0}1.4\%$} & \multirow{2}{*}{$\phantom{0}1.2\%$} & \multirow{2}{*}{$\phantom{0}0.01\%$} \\ 
 & & w/ spin & \multirow{2}{*}{$\phantom{0}3.7\%$} & \multirow{2}{*}{$\phantom{0}3.7\%$} & \multirow{2}{*}{$\phantom{0}0.79\%$} \\ 
 \cline{2-6} & \multirow{4}{*}{\makecell{\textsc{PDB} \\ 
 prior}} & w/o spin & \multirow{2}{*}{$\phantom{0}7.2\%$} & \multirow{2}{*}{$\phantom{0}6.4\%$} & \multirow{2}{*}{$\phantom{0}0.26\%$} \\ 
 & & w/ spin & \multirow{2}{*}{$16\%$} & \multirow{2}{*}{$16\%$} & \multirow{2}{*}{$\phantom{0}1.5\%$} \\ 
 \bottomrule
\end{tabular*}
\end{table}

\section{Model Selection and Determining Outliers} \label{sec:model_selection}

For some types of constraints, modeling uncertainties can cause ambiguity and diverging results, possibly leading to inaccurate predictions for the EOS. 
However, as already demonstrated in Sec.~\ref{sec:new_NICER}, established constraints on the EOS can be used to determine whether a new data point matches well with our expectations from other observations. 
Here, we use this approach to specify a reference for different mass-radius estimates of \mbox{PSR J0030+0451} and different measurements of symmetry energy parameters. 

%When a certain measurement provides a posterior $P(\text{EOS}|O)$ on the EOS candidates, we can compare that posterior to our expectation about the EOS from the established constraints introduced in Sec.~\ref{sec:EOS_constraints}.
Specifically, the posterior predictive distribution as defined in Eq.~\eqref{eq:posterior_predictive} allows us to compare the plausibility of different (potentially conflicting) observations $O_1$ and $O_2$. Interpreted this way, the posterior predictiveness ratio 
\begin{align}
    \mathcal{P}= P(O_1|d)/P(O_2|d)
\end{align}
can be interpreted as a Bayes factor of the observation $O_1$ vs. $O_2$. However, in contrast to the actual Bayes factor, the posterior predictiveness ratio compares different data points instead of different model prescriptions and is calculated not with respect to some agnostic prior but rather including information from the previous constraints in $d$.
An additional metric is the Bayesian coherence, defined as
\begin{align}
    \mathcal{C} = \frac{\sum_{\text{EOS}} \mathcal{L}(\text{EOS}|O_1) \pi(\text{EOS})}{\sum_{\text{EOS}} \mathcal{L}(\text{EOS}|O_1) P(\text{EOS}|d) }\,,
\end{align}
where $\pi(\text{EOS})$ is the EOS prior. 
This can be interpreted as the evidence ratio for the hypotheses $H_1$ vs. $H_0$, where $H_1$ claims that the single observation $O_1$ and the remaining constraints $d$ are not governed by the same EOS, while $H_0$ assumes that $O_1$ and $d$ are described by the same \text{EOS}~\cite{Koehn:2024set}.
We will use both metrics to quantify the preference of the established constraints for new data points. 

\subsection{Comparison of NICER results for PSR J0030+0451}

\begin{table*}
\centering
\caption{Posterior predictive ratios for the different $M$-$R$ inferences from the NICER observation of \mbox{PSR J0030+0451}. The ratio is given with respect to the original analysis of Ref.~\cite{Riley:2019yda}; i.e., the higher the value, the more plausible the original analysis. We also show the 95\% credible intervals on the inferred mass $M$ and radius $R$.}
\label{tab:NICER_comparison}
\renewcommand{\arraystretch}{1.5}
\begin{tabular}{>{\arraybackslash} p {0.8 cm}  >{\centering\arraybackslash} p { 2 cm}  >{\centering\arraybackslash} p { 1 cm}  >{\centering\arraybackslash} p { 1.6 cm}  >{\centering\arraybackslash} p { 1.6 cm}  >{\centering\arraybackslash} p { 1.3 cm} >{\centering\arraybackslash} p { 1.3 cm} >{\centering\arraybackslash} p { 1.3 cm} }
\toprule
\toprule
Ref. & Hot-spot configuration & XMM used & $M$ ($\msun$) & $R$ (km) &Set 1* & Set 2* & Set 3*\\
\midrule
\cite{Riley:2019yda} & ST+PST & $\cross$ & $1.34_{-0.27}^{+0.29}$ & $12.71_{-2.16}^{+2.15}$ & 1.0 & 1.0 & 1.0\\
\midrule
\multirow{8}{8.5pt}{\cite{Vinciguerra:2023qxq}} & PDT-U & $\cross$ & $1.41_{-0.34}^{+0.39}$ & $13.12_{-2.20}^{+2.42}$ & 1.23 & 1.15 & 1.18 \\ 
& ST-U & $\cross$ & $1.12_{-0.11}^{+0.29}$ & $10.53_{-1.48}^{+2.57}$ & 1.56 & 2.72 & 2.9 \\ 
& ST+PST & $\cross$ & $1.37_{-0.30}^{+0.32}$ & $13.11_{-2.47}^{+2.34}$ & 1.29 & 1.23 & 1.22 \\ 
& ST+PDT & $\cross$ & $1.20_{-0.18}^{+0.29}$ & $11.16_{-1.43}^{+1.85}$ & 0.84 & 1.43 & 1.52 \\ 
& ST-U & \checkmark & $1.88_{-0.53}^{+0.22}$ & $15.12_{-3.72}^{+0.84}$ & 8.38 & 8.37 & 9.24 \\ 
& ST+PST & \checkmark & $1.93_{-0.30}^{+0.17}$ & $15.23_{-2.25}^{+0.72}$ & 26.31 & 23.49 & 27.1 \\ 
& ST+PDT & \checkmark & $1.40_{-0.22}^{+0.26}$ & $11.71_{-1.63}^{+1.74}$ & 0.66 & 0.86 & 0.88 \\ 
& PDT-U & \checkmark & $1.70_{-0.38}^{+0.34}$ & $14.44_{-2.06}^{+1.40}$ & 7.44 & 5.46 & 6.24 \\ 
\bottomrule
\end{tabular}
\end{table*}
\begin{figure}
    \centering
    \includegraphics[width=\linewidth]{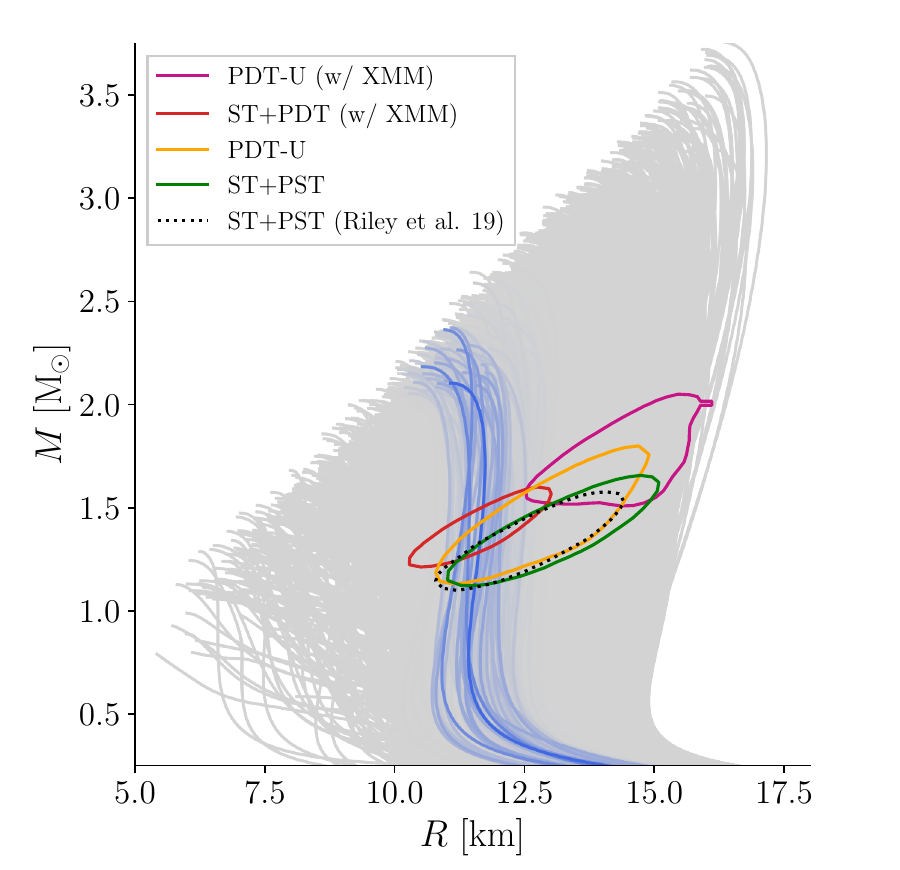}
    \caption{Mass-radius contours of \mbox{PSR J0030+0451} at 68\% credibility from the highest-evidence inferences of Ref.~\cite{Vinciguerra:2023qxq}. The dotted black contour shows the original posterior reported in Ref.~\cite{Riley:2019yda}. The background shows our set of EOS candidates color coded according to $P(\text{EOS}|\text{Set 1*})$, where light gray signifies low likelihood and blue signifies high likelihood.}
    \label{fig:NICER_comparison}
\end{figure}

Inferences of NS masses and radii from NICER observations require intricate modeling of the pulsed X-ray emission in combination with robust sampling techniques.
The pulsar \mbox{PSR J0030+0451} was observed with NICER, and the data was originally analyzed in Refs.~\cite{Riley:2019yda} and \cite{Miller:2019cac} to infer that the NS had a mass of $M\approx 1.35\,\msun$ and a radius of $R \approx 12.7$\, km.
Lately, new analyses of the raw data were reported in Ref.~\cite{Vinciguerra:2023qxq}, where the highest-evidence hot-spot model yielded $M=1.70^{+0.18}_{-0.19}\,\msun$ and $R=14.44^{+0.88}_{-1.05}$\, km at $68\%$ credibility. 
Ref.~\cite{Vinciguerra:2023qxq} differs from Ref.~\cite{Riley:2019yda} in updated sampling techniques, improved instrument response modeling, inclusion of the known XMM-Newton background, the assumption of no unknown component, and the chosen hot-spot models.
The authors in Ref.~\cite{Vinciguerra:2023qxq} emphasize that the new results were not run until full convergence and thus might not be considered robust.
Yet, we can compare these preliminary mass-radius estimates here and determine which of them appears more plausible given the other constraints from sets 1*, 2*, and 3*. The star signifies here that the original \mbox{NICER J0030+0451} measurement is removed for this purpose from the constraint sets.

In Fig.~\ref{fig:NICER_comparison}, we show the mass-radius contours from those hot-spot models with the highest evidence in Ref.~\citep{Vinciguerra:2023qxq} compared to our EOS posterior for set 1*. 
The labels for these different hot-spot models correspond to the convention of Ref.~\citep{Vinciguerra:2023qxq}.
While the compactness $M/R$ for each inference is very similar \cite{Luo:2024lbz}, the preferred mass and radius ranges vary substantially. 
Simple visual inspection confirms that the higher radii and mass values from the new analysis with a PDT-U hot-spot configuration are less plausible given the constraints from set 1*. 
In Table~\ref{tab:NICER_comparison} we list the posterior predictive ratios of the different \mbox{PSR J0030+0451} NICER mass-radius inferences with respect to the original analysis of Ref.~\citep{Riley:2019yda}. 
Those analyses that infer a high mass and radius for \mbox{PSR J0030+0451} are disfavored compared to the original analysis. 
Specifically, when XMM~data is included, the result of the PDT-U hot-spot model is $\gtrsim 5$ times less plausible than the original inference, despite having the highest evidence among these analyses. 
Moreover, the ST+PST and ST-U models are substantially disfavored against the original inference when the XMM data is used. 
This is also indicated by their coherence ratio, which across all constraint sets is $>7$ for the ST+PST model with XMM data, $>2.5$ for ST-U with XMM data, and $>3$ for PDT-U with XMM data.
For comparison, the original analysis of Ref.~\cite{Riley:2019yda} has a coherence ratio of roughly $0.6$.
Hence, if the inference of the NICER and XMM data with the ST+PST model was correct, it would require \mbox{PSR J0030+0451} to be described by a different EOS than the constraints in set 1, 2, and 3 with a Bayes factor of roughly $7$, whereas in case of the original analysis, the hypothesis that all constraints are governed by the same EOS is preferred.

All the other analyses from Ref.~\cite{Vinciguerra:2023qxq} are roughly equal in plausibility to the analysis from Ref.~\cite{Riley:2019yda}. 
Therefore, we propose that for the time being the original posterior from Ref.~\cite{Riley:2019yda} should be used for EOS inference, as it yields comparable results and the sampling there was run to convergence.

\subsection{Comparison for different measurements of the symmetry energy}
\begin{figure}
    \centering
    \includegraphics[width = \linewidth]{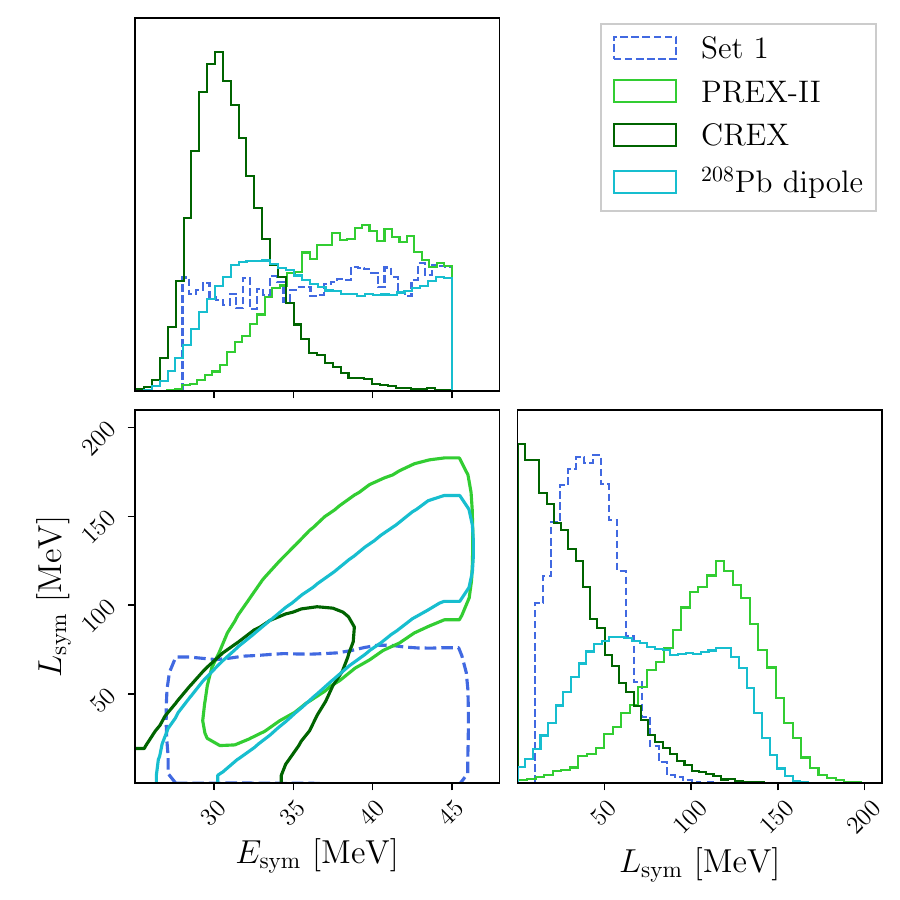}
    \caption{Posterior on the symmetry energy parameters $E_\sym$ and $L_\sym$ from different constraints. The contours show the 95\% credibility regions. Different constraints are color coded, and the dashed line refers to the constraint from set 1.}
    \label{fig:symmetry_energy}
\end{figure}
\begin{table}
\centering
\caption{Posterior predictive ratios of different symmetry energy measurements. The ratio is given with respect to CREX; i.e., the higher the value, the more plausible CREX is compared to the measurement in the row.}
\label{tab:symmetry_energy}
\begin{tabular}{>{\centering\arraybackslash} p { 4 cm}   >{\centering\arraybackslash} p { 1.3 cm} >{\centering\arraybackslash} p { 1.3 cm} >{\centering\arraybackslash} p { 1.3 cm} }
\toprule
\toprule
Measurement & Set 1 & Set 2 & Set 3\\
\midrule
CREX & 1.0 & 1.0 & 1.0 \\ 
PREX-II & 6.49 & 4.67 & 4.43 \\ 
$^{208}$Pb dipole & 2.03 & 1.70 & 1.66 \\ 
\bottomrule
\end{tabular}
\end{table}
The uncertainty in the neutron-star EOS around nuclear saturation density can be summarized by the poorly constrained parameters of the symmetry energy $e_{\text{sym}}(n)$. 
The latter is defined as the energy difference between pure neutron and symmetric nuclear matter and is a function of the number density $n$ through
\begin{eqnarray}
\label{eq:sym}
&&\hspace{-1cm}e_{\sym}(n) = E_\sym + L_\sym x + K_\sym \frac{x^2}{2}+ Q_\sym \frac{x^3}{3!} \nonumber \\ 
&&\hspace{2cm} + Z_\sym \frac{x^4}{4!} + \dots \, , \\
&&\hspace{0cm} x = \frac{n-n_\sat}{3n_\sat}\,. \nonumber
\end{eqnarray}
The symmetry energy $E_\sym$ and the symmetry energy slope $L_\sym$ at saturation density have been assessed by different nuclear experimental observations. 
Recently, a lot of attention has been paid to the CREX and PREX experimental campaigns at Jefferson Lab. 
PREX and CREX extracted $E_\sym$ and $L_\sym$ from measuring the parity-violating asymmetry in electron scattering with $^{48}$Ca~\cite{CREX:2022kgg} and $^{128}$Pb nuclei~\cite{PREX:2021umo}. 
A different set of measurements utilizes the electric dipole response and its relation to the symmetry energy~\cite{Tamii:2011pv, Roca-Maza:2015eza, Birkhan:2016qkr, Rossi:2013xha, Essick:2021ezp}, or neutron-skin measurements from strong probes, e.g., in Refs.~\cite{Giacalone:2023cet, Zenihiro:2018rmz, Friedman:2012pa, Tarbert:2013jze}. 
In Fig.~\ref{fig:symmetry_energy}, we show the posterior estimates on $E_\sym$ and $L_\sym$ from PREX-II, CREX, and the dipole measurement of $^{208}$Pb \citep{Tamii:2011pv}. 
These posteriors are taken from Ref.~\cite{Koehn:2024set} and employ a correlation between $E_\sym$ and $L_\sym$ established by an ensemble of energy density functionals.
It has been noted that the comparatively large $L_\sym$ from PREX-II is at odds with the smaller value from CREX \cite{Lattimer:2023rpe, Miyatsu:2023lki, Reed:2023cap, Mondal:2022cva}, though the two measurements can be reconciled when for instance analyzed with Skyrme functionals within uncertainties~\cite{Zhang:2022bni}.
Here, we compare the measurements of the symmetry energy from the CREX, PREX-II, and dipole polarization measurements and assess which data is more plausible given the remaining constraints on the EOS. 

For each of our EOS candidates, there are corresponding values for $E_\sym$ and $L_\sym$. Taking the measurement posterior from e.g., PREX-II, we can write the likelihood of the EOS as
\begin{align}
    \mathcal{L}(\text{EOS}|\text{PREX-II}) = P(E_\sym, L_\sym|\text{PREX-II})
\end{align}
where the arguments on the right-hand side are taken from the EOS. 
From there, we can again determine the posterior predictive ratios based on sets 1, 2, and 3 which are provided in Table~\ref{tab:symmetry_energy}. 
As expected, the lower value for $L_\sym$ of the CREX measurement is preferred over the large value by PREX-II with a factor of $\gtrsim 6$. 
The coherence ratio for the PREX-II measurement is 2.53, 2.13, and 2.18 for sets 1, 2, and 3 respectively.
This indicates again a mild preference of the PREX-II measurement for EOS candidates other than those favored by sets 1, 2, and 3. 
The electric dipole measurement of $^{208}$Pb mainly restricts a relationship between $E_\sym$ and $L_\sym$ and does not predict a single value. Therefore, its posterior is broad and there is no substantial preference when compared with PREX-II or CREX.

However, it is worth mentioning that the preference for lower $L_\sym$ in our constraint sets mainly arises from \chiEFT, since it essentially rules out any $L_\sym > 100$\,MeV. 
In fact, when we only use GW170817 as a constraint on our EOS set, the posterior predictiveness ratio of CREX vs. PREX-II drops to 1.76. 
It drops further to 2.16 when we additionally include the heavy pulsar mass limits.
Astrophysical observations of NS do not strongly constrain the symmetry parameters at nuclear saturation density, and hence, there is no preference for PREX-II or CREX based on GW170817 alone.

\section{Conclusion}
Existing knowledge about the EOS already allows for stringent limits on radii and maximum masses of NSs. 
It is based on combining a multitude of observational and theoretical data that cover different density regimes. 
In this work, we compared existing constraints to the new NICER observation of \mbox{PSR J0437-4715} and concluded that this new measurement agrees with hitherto expectations about the EOS. 
We combined this new information with previous constraints to receive an updated estimate of the EOS, finding $R_{1.4} = 12.05^{+0.84}_{-0.82}$\,km and $\mtov = {2.22}^{+0.38}_{-0.19}\,\msun$ (see Table~\ref{tab:new_NICER}).
This is consistent with other EOS analyses incorporating the new measurement~\cite{Rutherford:2024srk, Tang:2024jvs}.

The inferred knowledge about the EOS can be used to distinguish between BHs and NSs and to analyze aberrant data points concerning the EOS itself.
As an example of the former, we determined the nature of two compact objects in the posited lower mass gap, namely the primary object in GW230529 and the companion of \mbox{PSR J0514-4002E}. 
Based on our constraints of the EOS, we found that in both cases the object is likely not a NS. 
In particular, the probability for the primary of GW230529 being a NS are $\lesssim 16\%$, and $\lesssim 32\%$ for the companion of \mbox{PSR J0514-4002E}.
However, the exact likelihoods vary depending on which constraints are employed and on the details of how the masses are inferred from the measured data. 
In particular, the probability of either of these compact objects being a NS drops significantly when the remnant mass of GW170817 is used as an upper boundary on the Kepler limit, i.e., set 3 from Table~\ref{tab:sets}. 
Since other EOS constraints also imply a TOV mass of $\lesssim 2.5\,\msun$, it seems unlikely that any of the tentative mass gap objects with significant posterior support around \qty{3}{\msun} is a NS.

The current knowledge about the EOS is also sufficient to question some of the auxiliary NICER posteriors for \mbox{PSR J0030+0451} from Ref.~\cite{Vinciguerra:2023qxq}. 
The hot-spot configurations that predict higher masses and radii are noticeably disfavored compared to the original analysis of Ref.~\cite{Riley:2019yda}, with a posterior predictive ratio of $\gtrsim 5$. 
While this is not sufficient for immediate rejection (except maybe in case of ST+PST with XMM data, where the ratio exceeds 20), we conclude that the original analysis remains among the most plausible ones. 
Similarly, current EOS constraints prefer the CREX measurement of the symmetry energy over the one from PREX-II with a posterior predictive ratio of $\sim 6$, confirming the slight tension of the PREX-II results noted previously \cite{Miyatsu:2023lki, Reed:2023cap, Mondal:2022cva, Reinhard:2022inh}. 
Yet, this preference is mainly dominated by the inclusion of constraints from \chiEFT\ and is not substantial enough to make strong claims about PREX-II given the measurement uncertainties. 
This indicates the need for additional independent constraints of the EOS around nuclear saturation density \citep{Lattimer:2023rpe, Tsang:2023vhh, Becker:2018ggl} to effectively constrain the symmetry energy, as well as NS observations with higher precision to obtain better estimates of the EOS across several densities~\cite{Iacovelli:2023nbv, Branchesi:2023mws, Raithel:2019ejc, Essick:2021kjb}. 

Overall, we have demonstrated an application of the current EOS knowledge to compare additional and new data points. 
Significant ambiguity in such assessments will remain as long as the EOS constraints are not stringent enough and the uncertainty of the measurements under consideration remains broad. 

\begin{acknowledgments}

We thank Ewan D. Barr and Paulo C.C. Freire for fruitful discussions and for providing the data of the PSR J0514-4002E system. 
We thank Brendan T.\ Reed for discussions regarding the symmetry energy measurements.
We further want to express our gratitude to Olaf Drümmer and Peter Torsten Lübbe of Alter Pferdestall Fahren, whose hospitality drove the conceptual stages of this work.
We also thank the Institute for Nuclear Theory at the University of Washington for its kind hospitality and stimulating research environment. 
We acknowledge funding from the Daimler and Benz Foundation for the project “NUMANJI”. 
The project was also supported by the European Union (ERC, SMArt, 101076369). 
Views and opinions expressed are those of the authors only and do not necessarily reflect those of the European Union or the European Research Council. Neither the European Union nor the granting authority can be held responsible for them.
T.W. and P.T.H.P are supported by the research program of the Netherlands Organization for Scientific Research (NWO).
R.S. acknowledges support from the Nuclear Physics from Multi-Messenger Mergers (NP3M) Focused Research Hub which is funded by the National Science Foundation under Grant Number 21-16686, and by the Laboratory Directed Research and Development program of Los Alamos National Laboratory under project number 20220541ECR.
I.T. was supported by the U.S. Department of Energy, Office of Science, Office of Nuclear Physics, under contract No.~DE-AC52-06NA25396, by the Laboratory Directed Research and Development program of Los Alamos National Laboratory under project number 20230315ER, and by the U.S. Department of Energy, Office of Science, Office of Advanced Scientific Computing Research, Scientific Discovery through Advanced Computing (SciDAC) NUCLEI program.
This research was supported in part by the INT's U.S. Department of Energy grant No. DE-FG02-00ER41132.
This research used resources of the National Energy Research Scientific Computing Center (NERSC), a U.S. Department of Energy Office of Science User Facility located at Lawrence Berkeley National Laboratory, operated under Contract No. DE-AC02-05CH11231 using NERSC award ERCAP0028885.
This research used resources of the national supercomputer HPE Apollo Hawk at the High Performance Computing (HPC) Center Stuttgart (HLRS) under the grant number GWanalysis/44189.
This research used resources of the the GCS Supercomputer SuperMUC\_NG at the Leibniz Supercomputing Centre (LRZ) [project pn29ba].

This material is based upon work supported by NSF's LIGO Laboratory which is a major facility fully funded by the National Science Foundation.
\end{acknowledgments}

\appendix*
\section{Excluding \mbox{PSR J0437-4715} from the constraint sets}
\label{app:wo_new_NICER}
In this Appendix, we present how our results change when we exclude the $M$-$R$ measurement of \mbox{PSR J0437-4715} in set 1, set 2, and set 3.
In Tables~\ref{tab: classification results app}, \ref{tab:NICER_comparison app}, and \ref{tab:symmetry_energy_app}, we show the adapted results for the compact object classification, hot-spot model preference, and symmetry energy measurements.
Since the impact on the TOV mass is small, the classification for the compact objects accompanying \mbox{PSR J0514-4002E} and GW230529 does not change much compared to the original Table~\ref{tab: classification results}.
For the different hot-spot models of Ref.~\citep{Vinciguerra:2023qxq}, the NICER observation of \mbox{PSR J0437-4715} indeed increased the rejection of the ST+PST and PDT-U inference with XMM data notably, especially for set 1. 
The preference or disfavor of the remaining hot-spot models is only marginally affected.
\mbox{PSR J0437-4715} measurement also increases the preference for CREX over PREX-II, as it allows for softer EOSs, though the difference in the posterior predictive ratios is only small.

\begin{table}
\renewcommand{\arraystretch}{1.25}
\centering
\caption{Probability that the companion of \mbox{PSR J0514-4002E} and the primary component of GW230529 are NSs. This is the equivalent for Table~\ref{tab: classification results}.}
\label{tab: classification results app}
\begin{tabular*}{0.975\linewidth}{@{\extracolsep{\fill}} p {1.8 cm} p{1.2 cm} >{\centering\arraybackslash}p {1 cm}  c c c}
\toprule\toprule
Object & & & Set $1$ & Set $2$ & Set $3$ \\
\midrule 
\multirow{3}{*}{\makecell{PSR J0514\\-4002E}} & \multicolumn{2}{l}{$M_2>\qty{1.17}{\msun}$}& $35\%$ & $33\%$ & $\phantom{0}2.3\%$ \\ 
 & \multicolumn{2}{l}{$M_2$ unconstrained} & $27\%$ & $25\%$ & $\phantom{0}1.4\%$ \\ 
 & \multicolumn{2}{l}{$M_2\in P^{\rm gal}(M_{\rm PSR})$} & $17\%$ & $\phantom{0}7.9\%$ & $\phantom{0}1.3\%$ \\ 
 \midrule \multirow{8}{*}{GW230529} & \multirow{4}{*}{\makecell{default \\ prior}} & w/o spin & \multirow{2}{*}{$\phantom{0}1.9\%$} & \multirow{2}{*}{$\phantom{0}1.2\%$} & \multirow{2}{*}{$\phantom{0}0.01\%$} \\ 
 & & w/ spin & \multirow{2}{*}{$\phantom{0}4.3\%$} & \multirow{2}{*}{$\phantom{0}3.7\%$} & \multirow{2}{*}{$\phantom{0}0.79\%$} \\ 
 \cline{2-6} & \multirow{4}{*}{\makecell{\textsc{PDB} \\ 
 prior}} & w/o spin & \multirow{2}{*}{$\phantom{0}9.4\%$} & \multirow{2}{*}{$\phantom{0}6.5\%$} & \multirow{2}{*}{$\phantom{0}0.27\%$} \\ 
 & & w/ spin & \multirow{2}{*}{$19\%$} & \multirow{2}{*}{$16\%$} & \multirow{2}{*}{$\phantom{0}1.5\%$} \\ 
 \bottomrule
\end{tabular*}
\end{table}

\begin{table}
\centering
\caption{Posterior predictive ratios for the different $M$-$R$ inferences from the NICER observation of \mbox{PSR J0030+0451}. This corresponds to Table~\ref{tab:NICER_comparison}.}
\label{tab:NICER_comparison app}
\renewcommand{\arraystretch}{1.5}
\begin{tabular}{>{\arraybackslash} p {0.8 cm}  >{\centering\arraybackslash} p { 2 cm}  >{\centering\arraybackslash} p { 1 cm}  >{\centering\arraybackslash} p { 1.3 cm} >{\centering\arraybackslash} p { 1.3 cm} >{\centering\arraybackslash} p { 1.3 cm} }
\toprule
\toprule
Ref. & Hot-spot configuration & XMM used & Set 1* & Set 2* & Set 3*\\
\midrule
\cite{Riley:2019yda} & ST+PST & $\cross$ & 1.0 & 1.0 & 1.0\\
\midrule
\multirow{8}{8.5pt}{\cite{Vinciguerra:2023qxq}} & PDT-U & $\cross$ & 1.19 & 1.15 & 1.17 \\ 
& ST-U & $\cross$ & 2.06 & 3.02 & 3.26 \\ 
& ST+PST & $\cross$ & 1.25 & 1.22 & 1.21 \\ 
& ST+PDT & $\cross$ & 1.11 & 1.61 & 1.74 \\ 
& ST-U & \checkmark & 7.8 & 8.16 & 8.96 \\ 
& ST+PST & \checkmark & 21.6 & 22.29 & 25.42 \\ 
& ST+PDT & \checkmark & 0.79 & 0.93 & 0.95 \\ 
& PDT-U & \checkmark & 5.36 & 4.99 & 5.57 \\ 
\bottomrule
\end{tabular} % TODO: UPDATE
\end{table}

\begin{table}
\centering
\caption{Posterior predictive ratios of different symmetry energy measurements. This corresponds to Table~\ref{tab:symmetry_energy}.}
\label{tab:symmetry_energy_app}
\begin{tabular}{>{\centering\arraybackslash} p { 4 cm}   >{\centering\arraybackslash} p { 1.3 cm} >{\centering\arraybackslash} p { 1.3 cm} >{\centering\arraybackslash} p { 1.3 cm} }
\toprule
\toprule
Measurement & Set 1 & Set 2 & Set 3\\
\midrule
CREX & 1.0 & 1.0 & 1.0 \\ 
PREX-II & 6.15 & 4.54 & 4.25 \\ 
$^{208}$Pb dipole & 1.97 & 1.67 & 1.62 \\ 
\bottomrule
\end{tabular}
\end{table} 

\vfill
\newpage

\begin{acronym}
    \acro{NICER}[NICER]{Neutron Star Interior Composition Explorer}
    \acro{ASD}[ASD]{amplitude spectral density}
    \acro{BBH}[BBH]{binary black hole}
    \acro{BH}[BH]{black hole}
    \acrodefplural{BHs}{black holes}
    \acro{BNS}[BNS]{binary neutron star}
    \acro{CBC}[CBC]{compact binary coalescences}
    \acro{CE}[CE]{Cosmic Explorer}
    \acro{EOB}[EOB]{effective one-body}
    \acro{EM}[EM]{electromagnetic}
    \acro{EOS}[EOS]{equation of state}
    \acro{ET}[ET]{Einstein Telescope}
    \acro{GW}[GW]{gravitational-wave}
    \acrodefplural{GWs}{gravitational waves}
    \acro{HF}[HF]{high-frequency}
    \acro{IMRPhenom}[\texttt{IMRPhenom}]{inspiral-merger-ringdown phenomenological}
    \acro{LF}[LF]{low-frequency}
    \acro{LMXB}{low-mass X-ray binary}
    \acrodefplural{LMXB}{low-mass X-ray binaries}
    \acro{MSP}[MSP]{millisecond pulsar}
    \acro{NS}[NS]{neutron star}
    \acrodefplural{NSs}{neutron stars}
    \acro{NSBH}[NSBH]{neutron star-black hole}
    \acro{PN}[PN]{post-Newtonian}
    \acro{PSD}[PSD]{power spectral density}
    \acro{SNR}[SNR]{signal-to-noise ratio}
    \acro{TOV}[TOV]{Tolman-Oppenheimer-Volkoff}

\end{acronym}



\bibliography{references}{}
\bibliographystyle{apsrev4-1}

\end{document}